\documentclass[twocolumn]{aastex63}
\usepackage{amsmath}
\usepackage{hyperref}
\usepackage{soul}

\begin{document}

\title{ \emph{Fermi}-LAT detection of extended gamma-ray emission in the vicinity of SNR G045.7$-$00.4: evidence for escaping cosmic rays interacting with the surrounding molecular clouds}

\author[0000-0001-6863-5369]{Hai-Ming Zhang}
\affil{School of Astronomy and Space Science, Nanjing University, Nanjing 210023, Jiangsu, China; xywang@nju.edu.cn}
\affil{Key laboratory of Modern Astronomy and Astrophysics(Nanjing University), Ministry of Education, Nanjing 210023, People Republic of China}
\author[0000-0003-1576-0961]{Ruo-Yu Liu}
\affil{School of Astronomy and Space Science, Nanjing University, Nanjing 210023, Jiangsu, China; xywang@nju.edu.cn}
\affil{Key laboratory of Modern Astronomy and Astrophysics(Nanjing University), Ministry of Education, Nanjing 210023, People Republic of China}
\author{Yang Su}
\affil{Purple Mountain Observatory and Key Laboratory of Radio Astronomy, Chinese Academy of Sciences, Nanjing 210023, People Republic of China}
\author{Hui Zhu}
\affil{National Astronomical Observatories, Chinese Academy of Sciences, CAS, Jia-20 Datun Road, Chaoyang District, Beijing 100012 , China}
\author{Shao-Qiang Xi}
\affil{Key Laboratory of Particle Astrophyics \& Experimental Physics Division \& Computing Center, Institute of High Energy Physics, Chinese Academy of Sciences, Beijing 100049 , China}
\author[0000-0002-5881-335X]{Xiang-Yu Wang}
\affil{School of Astronomy and Space Science, Nanjing University, Nanjing 210023, Jiangsu, China; xywang@nju.edu.cn}
\affil{Key laboratory of Modern Astronomy and Astrophysics(Nanjing University), Ministry of Education, Nanjing 210023, People Republic of China}

\begin{abstract}

We present the analysis of \emph{Fermi} Large Area Telecope (LAT) data of the gamma-ray emission in the vicinity of a radio supernova remnant (SNR), G045.7$-$00.4. To study the origin of the gamma-ray emission, we also make use of the CO survey data of Milky Way Imaging Scroll Painting to study the massive molecular gas complex that surrounds the SNR.  The whole size of the  GeV emission is significantly larger than that of the radio morphology. Above 3 GeV, the GeV emission is resolved into two sources: one  is spatially consistent with the position of the SNR with a size comparable to that of the radio emission, and the other is located outside of the western boundary of the SNR and spatially coincident with the densest region of the surrounding molecular cloud. We suggest that the  GeV emission of the western source may arise from  cosmic rays (CRs) which have escaped the SNR and illuminated the surrounding molecular cloud.  We find that the  gamma-ray spectra of the western source  can be consistently explained by this scenario with a total energy of   $\sim 10^{50}{\rm erg}$ in escaping CRs assuming the escape is isotropic.

\end{abstract}


\section{Introduction}           
\label{sect:intro}

Diffusive shock acceleration (DSA) operating at expanding shock waves of supernova remnants  is widely
believed to be the mechanism  converting the kinetic energy released by supernova explosions into the energy of cosmic rays (e.g., \citealt{Malkov2001}).
In the DSA theory, cosmic rays (CRs) being accelerated at shocks must be scattered by self-generated magnetic turbulence. Since the highest-energy CRs in the shock precursor are prone to lack self-generated turbulence, they are expected to escape the shock. The DSA theory generally predicts that a substantial fraction of
the shock energy is carried away by escaping CRs. In the presence of  molecular clouds surrounding the supernova remnant (SNR),  escaping CRs can illuminate the clouds through $pp$ interactions, producing  gamma-ray emission with a flux depending on the amount of nuclear CRs released by a
supernova remnant and the diffusion coefficient in
the interstellar medium \citep{Aharonian1996,Rodriguez Marrero2008,Aharonian2004,Gabici2009}. H.E.S.S. observations reveal  a complex of sources (HESS J1800-240A, B and C) $\sim0.5\degr$
south of the SNR W28, coincident with
molecular clouds in the field, and the Large Area Telescope (LAT) on board the \emph{Fermi} satellite  reveals a similar structure in GeV energies \citep{Abdo2010,Hanabata2014}.  The GeV-TeV gamma-ray emission around W28
can be regarded as a realization of this scenario \citep{Aharonian2008a}.  Another example is the detection of two extended gamma-ray structures located at two opposite edges of the  SNR W44 by the \textit{Fermi}-LAT \citep{Uchiyama2012,Peron2020}. The gamma-ray emission coincides with the molecular cloud complex that surrounds SNR W44.
The gamma-ray emission that appears to come from the surrounding molecular cloud complex can be ascribed to the cosmic rays that have escaped from W44. The total kinetic energy channeled into the escaping CRs are estimated to be
larger than a few $10^{49}{\rm \ erg}$ in both W28 and W44, although the exact number depends on the value of the diffusion coefficient of escaping CRs.

\begin{figure*}
\includegraphics[angle=0,scale=0.5]{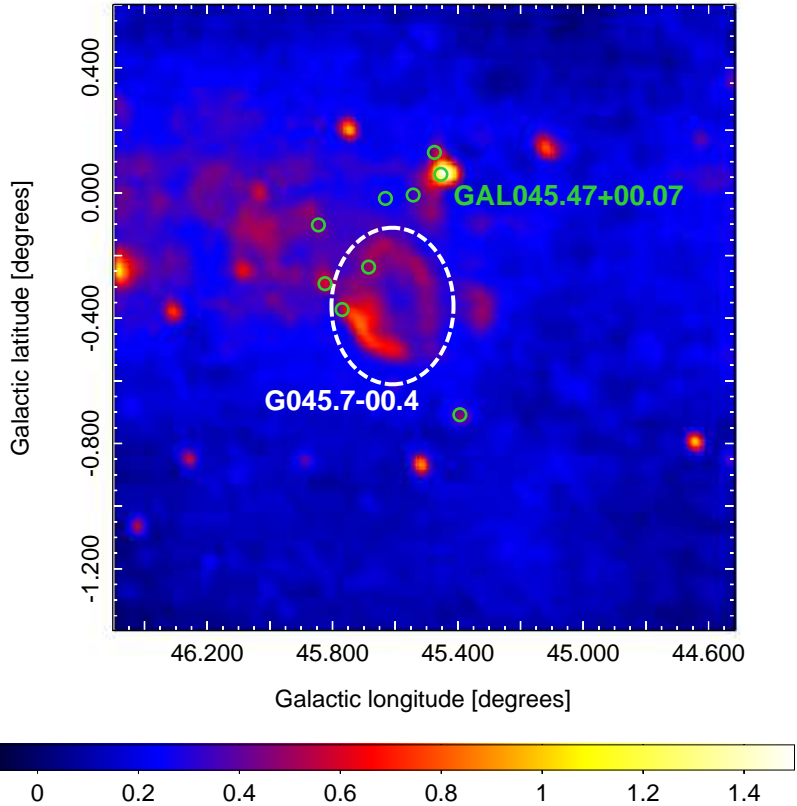}
\caption{Radio observation of the region around G045.7$-$00.4 at 200 MHz \citep{Hurley2019}. The white dashed ellipse shows the shell-like morphology of G045.7-00.4. The green circles represent the HII regions and the brightest HII-region is GAL 045.47+00.07 \citep{Bayandina2015}, as indicated in the plot.
The color bar is in unit $\rm Jy \ beam^{-1}$.
}
\label{rxmap}
\end{figure*}

SNR G045.7$-$00.4 is one of the 30 sources  classified
as likely GeV SNRs  in the first \emph{Fermi}-LAT supernova remnant catalog (with 36 months data in 1-100 GeV),  based on the spatial overlap of  sources detected at GeV energies with SNRs known from radio surveys \citep{Acero2016}. No significant extension is found with 36 months \emph{Fermi}-LAT  data \citep{Acero2016}.
In this paper, we report the  analysis result of the GeV gamma-ray emission from the direction of SNR G045.7$-$00.4 using over 12 years \emph{Fermi}-LAT data in 3-500 GeV energy band, and present the  CO observational results of  molecular clouds in this region. G045.7$-$00.4 is a shell-type SNR  with a radius of about $11'$ in the radio band \citep{Fuerst1987,Green2019}. Figure \ref{rxmap} shows the radio continuum emission map at 200 MHz \citep{Hurley2019}.  The spectrum from 1 to 10 GHz shows a power-law form of $S_\nu \propto \nu^{-0.3}$.  We will show that the GeV emission is much more extended than the radio SNR. Interestingly, CO observations reveal a giant molecular cloud complex surrounding SNR G045.7$-$00.4. This makes this source a likely analogy of  W28 and W44, with gamma-ray emission produced by escaping CRs illuminating the molecular cloud.

The paper is organized as follows. We first report the \emph{Fermi}-LAT observations of the region around SNR G045.7$-$00.4 in \S 2. The observation results of CO distribution of this region are reported in \S 3. Then we discuss the interpretation of the GeV sources in \S 4. Finally, we give discussions and conclusions.

\section{\emph{Fermi}/LAT Data Analysis}

The \textit{Fermi}-LAT is sensitive to $\gamma$-rays with energies from 20 MeV to over 300 GeV, and it has continuously monitored the sky since 2008 \citep{Atwood2009}. The  Pass 8 data taken from 2008 August 4 to 2020 November 17 are used to study the GeV emission around SNR G045.7$-$00.4 region.
The event class P8R3$\_$SOURCE (``evclass=128'')  and event type FRONT + BACK (``evtpye=3'' ) are used. The event class is the recommended class by the LAT team and provides good sensitivity for analysis of point sources and moderately extended sources\footnote{\url{https://fermi.gsfc.nasa.gov/ssc/data/analysis/documentation/Cicerone/Cicerone_Data/LAT_DP.html\#PhotonClassification}}.
We only consider the $\gamma$-ray events in the $1-500 \, \rm GeV$ energy range, with the standard data quality selection criteria ``$(DATA\_QUAL > 0)  \&\& (LAT\_CONFIG == 1)$". 
To minimize the contamination from the Earth limb, the maximum zenith angle is set to be 90$\degr$.
In this work, the publicly available software \textit{Fermitools} (ver. 1.2.23) is used to preform the data analysis.

\begin{figure*}
\includegraphics[angle=0,scale=0.38]{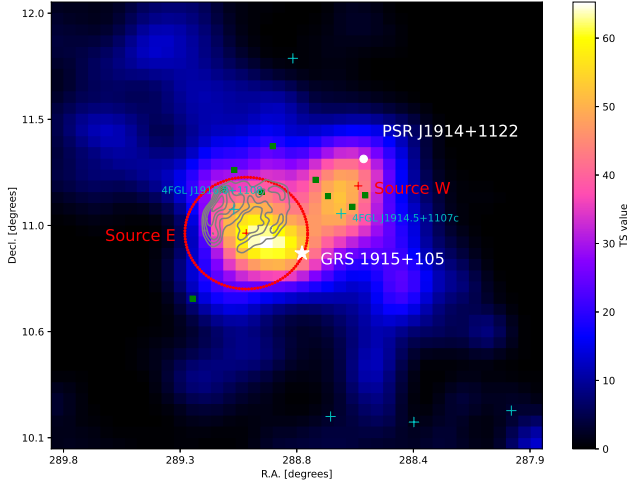}
\caption{$2\degr \times 2\degr$ TS map of the gamma-ray emission above 3 GeV measured by Fermi/LAT  around SNR G045.7$-$00.4. The red dashed circle shows the 68\% containment radius of Source E in the best-fit uniform disk model.
The grey contours show radio emission intensity of SNR G045.7$-$00.4 \citep{Hurley2019}. 
The cyan crosses represent the 4FGL sources. 
The green squares represent the HII regions. The white dot represents the position of PSR J1914+1122 \citep{Manchester2005}.
}
\label{Gevmap1}
\end{figure*}

Only the data within a $14\degr \times14\degr$ region of interest (ROI) centered on the position of G045.7$-$00.4 are considered for the binned maximum likelihood analysis.
The instrument response functions (IRFs) (\textit{$P8R3\_SOURCE\_V3$}) is used.
We include the diffuse Galactic interstellar emission (IEM, $gll\_iem\_v07.fits$),  isotropic emission (``$iso\_P8R3\_SOURCE\_V3\_v1.txt$'' ) and all sources listed in the fourth \textit{Fermi}-LAT catalog\citep{Abdollahi2020a} in the background model.
All sources within $4\degr$ of the center are set free.
The parameters of IEM and isotropic emission are also left free.
The maximum likelihood test statistic (TS) is used to estimate the significance of $\gamma$-ray sources, which is defined by TS$= 2 (\ln\mathcal{L}_{1}-\ln\mathcal{L}_{0})$, where $\mathcal{L}_{1}$ and $\mathcal{L}_{0}$ are maximum likelihood values for the background with target source and without the target source (null hypothesis).
Figure \ref{Gevmap1} shows the 3-500 GeV TS map in the the vicinity of G045.7$-$00.4 with the binned likelihood method provided by \textit{Fermitools}. The TS map is generated by only considering background fitting but not including 4FGL J1916.3+1108 and 4FGL J1914.5+1107c. 
An obvious excess of GeV emission is seen in the vicinity of G045.7$-$00.4. The excess appears to consist of two  parts, which are roughly consistent with two 4FGL sources in position (4FGL J1916.3+1108 and 4FGL J1914.5+1107c).
4FGL J1916.3+1108 is spatially coincident with SNR G045.7$-$00.4, while  4FGL J1914.5+1107c does not have any clear counterparts at other wavelengths.
Hereafter, we refer to the eastern part of the GeV excess as Source E and the western part as Source W.

\subsection{Morphological analysis}

\begin{figure*}
\includegraphics[angle=0,scale=0.25]{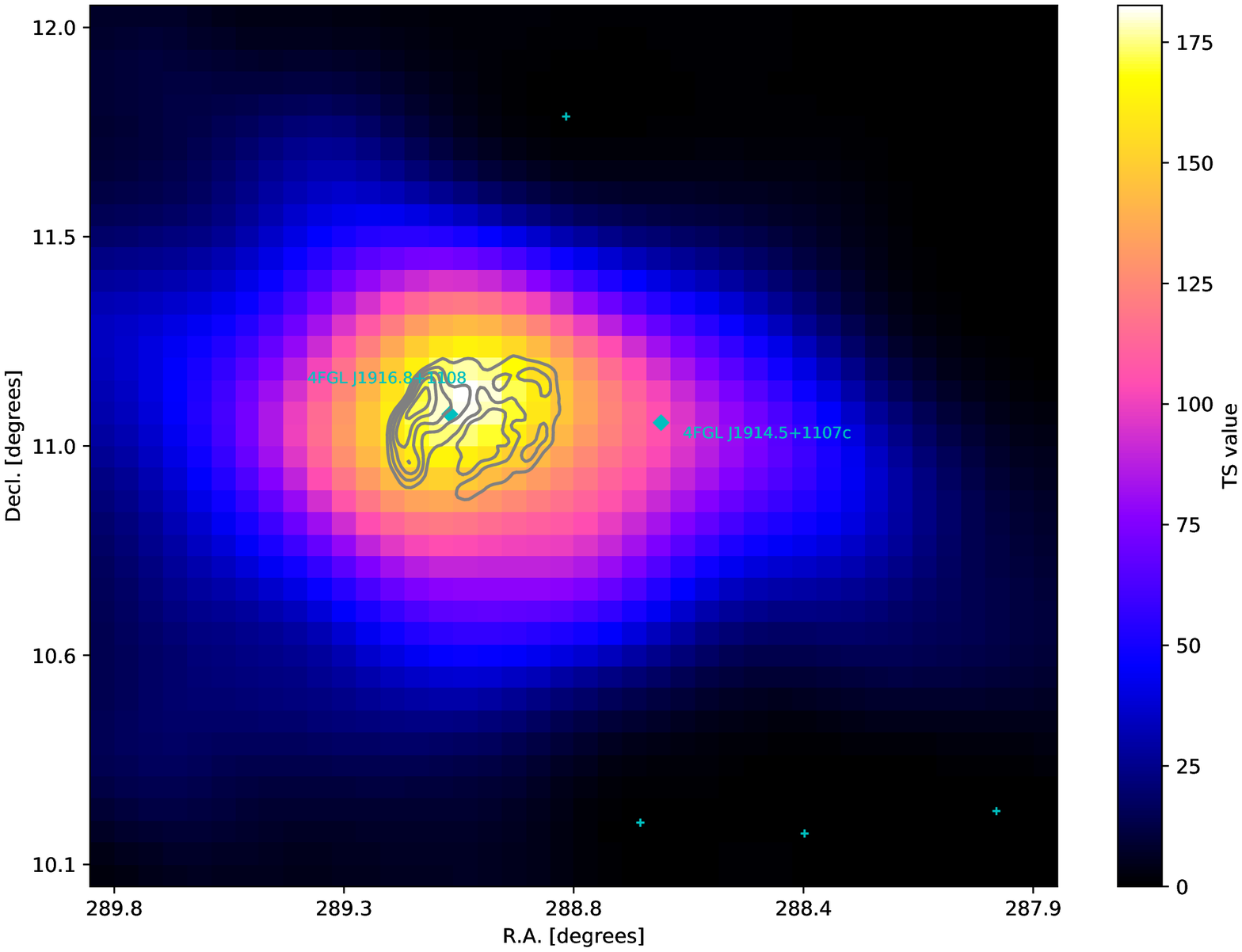}
\includegraphics[angle=0,scale=0.25]{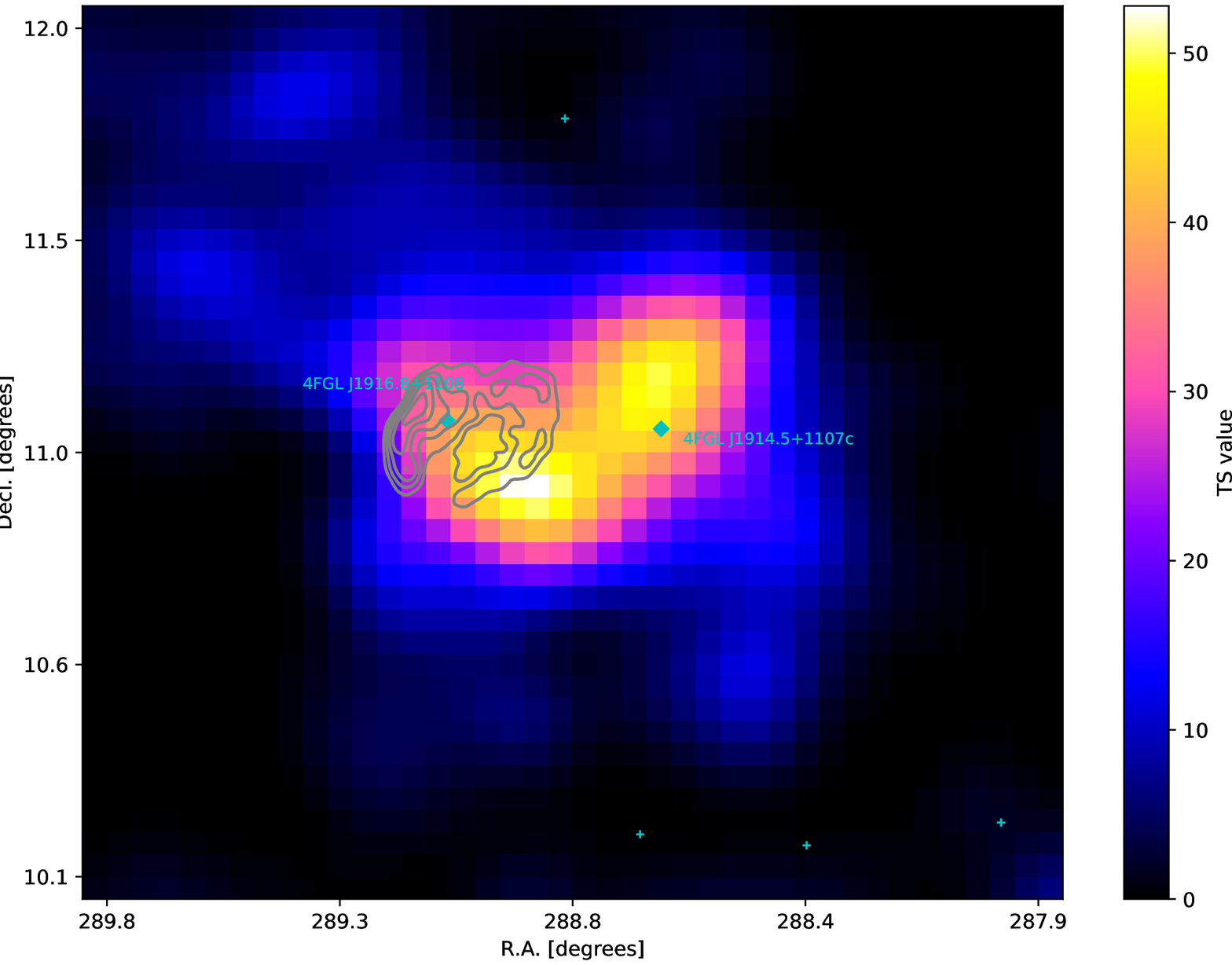}
\includegraphics[angle=0,scale=0.25]{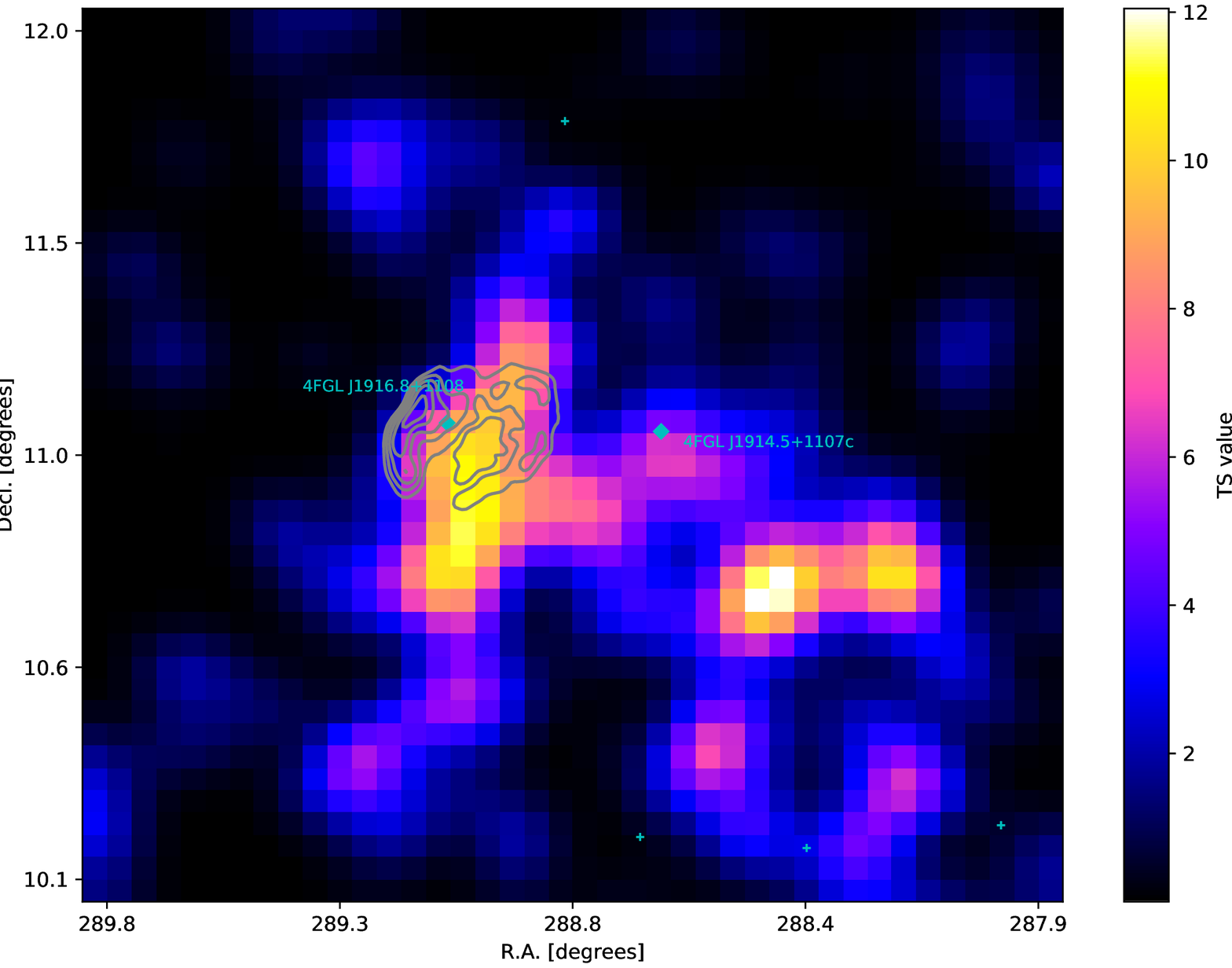}
\caption{$2\degr \times 2\degr$ TS map of the gamma-ray emission in 1-3 GeV (left plane), 3-10 GeV (middle plane) and 10-30 GeV (right plane) measured by \emph{Fermi}/LAT around SNR G045.7$-$00.4.
As in Figure \ref{Gevmap1}, the grey contours show radio emission intensity of SNR G045.7$-$00.4 \citep{Hurley2019}. The cyan crosses and diamonds represent the 4FGL sources.
}
\label{Gevbin}
\end{figure*}

We first study the energy-dependence of the morphology of the gamma-ray emission in the the vicinity of G045.7$-$00.4.
The TS map of the gamma-ray emission in 1-3 GeV, 3-10 GeV and 10-30 GeV are shown in the left, middle and right panel of Figure \ref{Gevbin}.
Similar to the TS map in 3-500 GeV, the TS map of the gamma-ray emission in 3-10 GeV shows two regions of gamma-ray excess: one is spatially consistent with the position of the radio SNR, and the other is on the west  of the radio remnant. In 10-30 GeV, only a marginal detection of gamma-ray emission at the position of the radio SNR is seen.
Taking into account of a better point-spread function (PSF) in the higher energy band as well as sufficient statistics for the analysis, we  considered events only above 3 GeV in the following morphological analysis.

We use the \emph{Fermipy} tool to quantitatively evaluate the extension and location of these two sources.
The uniform disk model is used to evaluate the extension of them. For the Source W,  we do not find significant extension. Under the assumption of a disk shape, the upper limit extension amounts to 21 arcmin at the 95\% confidence level. We thus treat Source W as a point-like source hereafter. The best-fit position of Source W is estimated to be ($\rm R.A.,Decl.$)=($288.57\degr\pm0.04\degr$,$11.25\degr\pm0.09\degr$) in the energy band above 3 GeV. This position is 8 arcmin away from the position of 4FGL J1914.5+1107c.

For the Source E, its extension in the uniform disk model is $\sigma=0.25\degr_{-0.05\degr}^{+0.08\degr}$ with a $\rm TS_{ext}=25.8$. The extension $\sigma$ represents the radius containing 68\% of the intensity, and we define it as the source size.
The $\rm TS_{ext}$ is defined as $\rm {TS_{ext}}=2(\ln\mathcal{L}_{ext}-\ln\mathcal{L}_{ps})$,
where $\mathcal{L}_{ext}$ is the maximum likelihood value for the extend model and $\mathcal{L}_{ps}$ is the maximum likelihood value for the point-like model.
Compared to the point-like model, the uniform disk model has one additional degree of freedom, and the extension significance is approximate to $\rm \sqrt{TS_{ext}} \ \sigma$ \citep{Abdollahi2020a,Abdollahi2020b}.
The values of $\rm TS_{ext}$ for the uniform disk model is 25.8, which rejects a point-like source hypothesis at $5.1\sigma$.
Also, we test a two-dimensional Gaussian model and find the results are quite similar, with $\sigma=0.24\degr_{-0.06\degr}^{+0.07\degr}$  and $\rm TS_{ext}=29.1(5.4\sigma)$.
The best-fit position for Source E is ($\rm R.A.,Decl.$)=($289.03\degr\pm0.02\degr$,$11.04\degr\pm0.03\degr$),
which is 7 arcmin away from the position of 4FGL J1916.3+1108.
The results of the morphology analysis are summarized in Table \ref{tab:morph}.

\subsection{Energy Spectrum}

\begin{figure*}
\includegraphics[angle=0,scale=0.45]{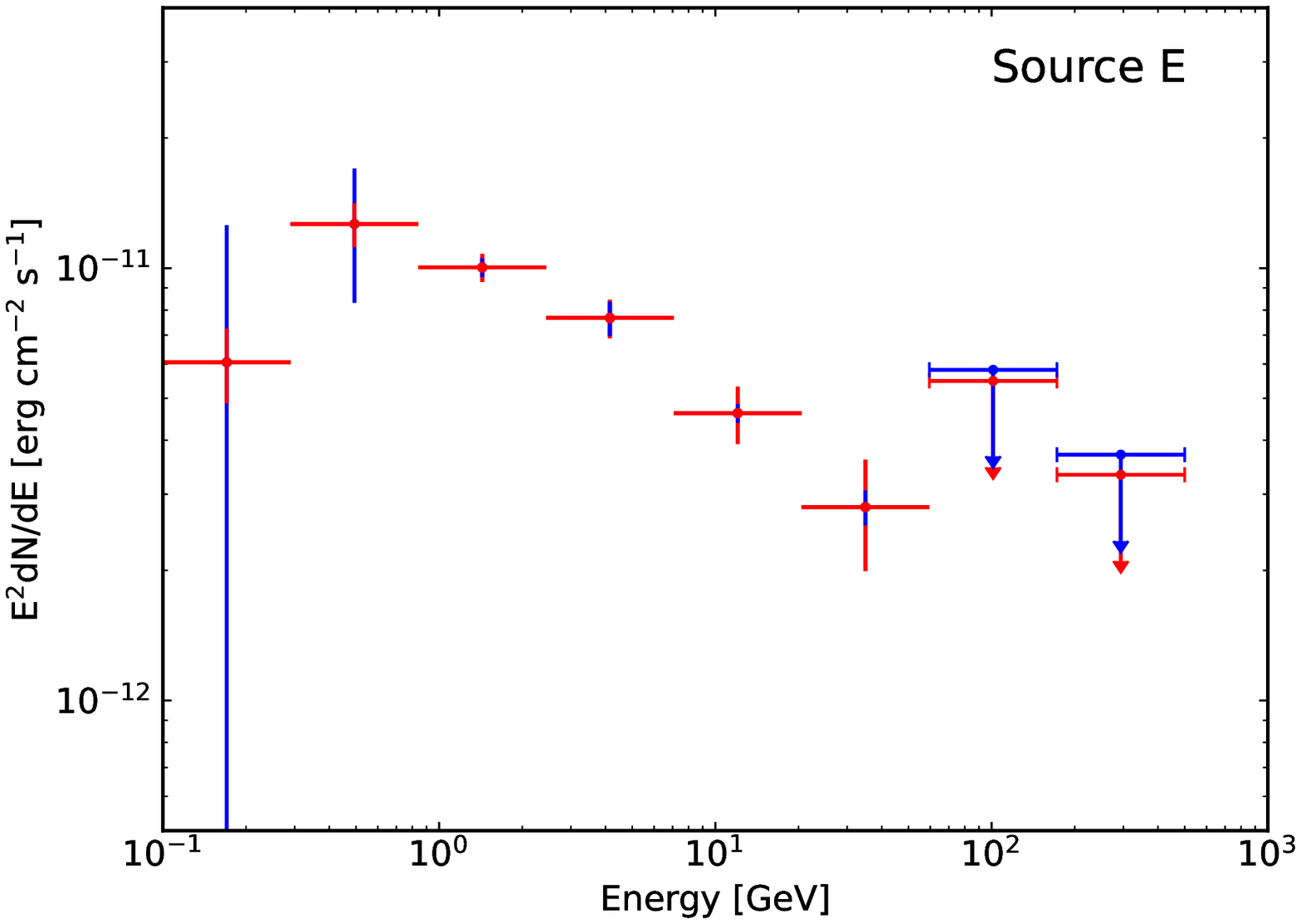}
\includegraphics[angle=0,scale=0.45]{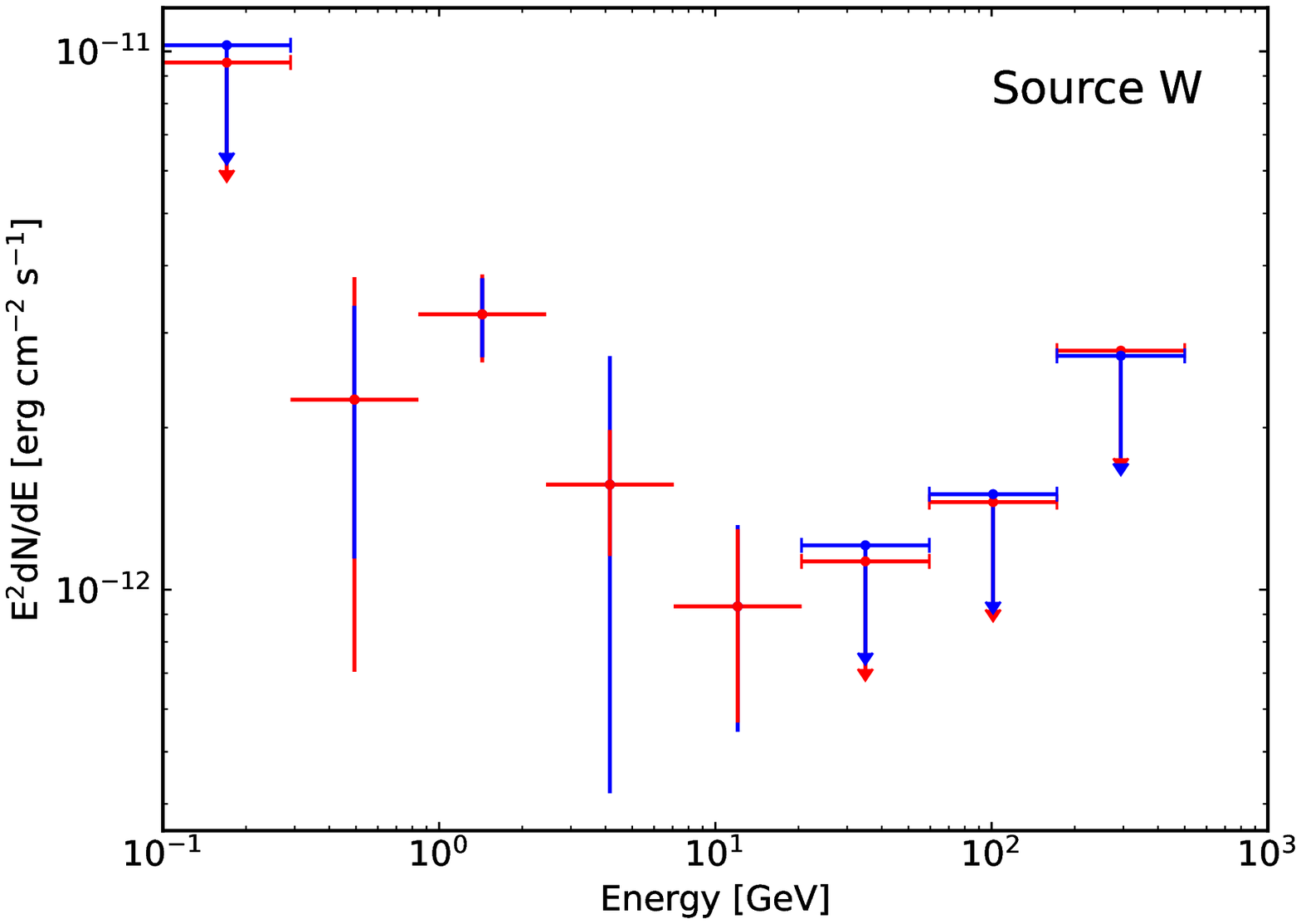}
\caption{SED of Source E (left plane) and Source W (right plane) as measured by \emph{Fermi}-LAT. Statistical and systematic uncertainties are given in red and blue, respectively. When the TS value of the data point is less than 4, an upper limit is calculated at 95\% confidence level using a Bayesian method. The blue upper limits show the limits due to the systematic effects, which are taken from the larger of the errors caused by the Galactic diffuse emission model and source spatial model.
}
\label{sedew}
\end{figure*}

After the morphology was fixed, we derive the gamma-ray spectral energy distribution (SED) of Source E and Source W. The results are shown in Figure \ref{sedew}.
When the TS value of spectra data point is less than 4, the upper limit is calculated at 95\% confidence level using a Bayesian method \citep{Helene1983}.
In this work, we  consider two  sources of systematic errors: (1) uncertainties due to imperfect modeling of the Galactic diffuse emission, $\sigma_{\rm IEM}$; (2) uncertainties due to the source spatial model, $\sigma_{\rm model}$.  $\sigma_{\rm IEM}$ are evaluated by going over the whole process using an alternative Galactic diffuse emission model ($gll\_iem\_v06.fits$). For the morphology analysis, as shown in Table \ref{tab:morph}, both the uniform disk  plus a point-like source model (Disk + PS) and the two-dimensional Gaussian plus a point-like source model (Gaussian + PS) can describe well the GeV morphology. Therefore,  $\sigma_{\rm model}$ is estimated by comparing the spatial model of Disk + PS with that of Gaussian + PS.
The total systematic errors ($\sigma_{sys}$), shown as blue error bars in Figure \ref{sedew}, are evaluated by adding in quadrature the uncertainties due to the Galactic diffuse model and the source spatial model as $\sigma_{sys}=\sqrt{\sigma_{\rm IEM}^{2}+\sigma_{\rm model}^{2}}$ \citep{Abdollahi2020b}.
In the energy band greater than 3 GeV, the spectra of both sources are well described by a power-law function with $\rm TS_{curve}<9$ ($\rm TS_{curve}=9$ corresponding to $3\sigma$ \citep{Abdollahi2020a}), where $\rm TS_{curve}$ is defined as $\rm{TS_{curve}}=2(\ln\mathcal{L}_{curved \  spectrum}-\ln\mathcal{L}_{power-law})$.
In this work, the curved spectrum represents a log parabola spectral shape.
But in the $>$ 0.1 GeV energy band, the spectrum of the Source E is well described by a log parabola function with $\rm TS_{curve}=68.34$ ($8.3\sigma$), while the spectrum of the Source W is still described by a power-law function with $\rm TS_{curve}<9$.
Using the alternative Galactic diffuse emission model, we find that the  $\rm TS_{curve}$ values for Source E and Source W are 28.06 ($5.3\sigma$) and 6.39 ($2.5\sigma$),  respectively.
Therefore, the spectral models are assumed to be a log parabola function for Source E and a power-law function for Source W above 0.1 GeV in this work. The spectral indices of the Source E and Source W are $-2.48\pm0.10_{stat}\pm0.05_{sys}$ and $-2.28\pm0.09_{stat}\pm0.08_{sys}$, respectively. The energy flux of Source E and Source W are $\rm (4.03\pm0.78_{stat}\pm0.64_{sys})\times 10^{-11} \ erg\ cm^{-2} \ s^{-1}$ and $\rm (1.11\pm0.22_{stat}\pm0.08_{sys})\times 10^{-11} \ erg\ cm^{-2} \ s^{-1}$  in the 0.1-500 GeV energy band, respectively.

\section{CO observations}

\begin{figure*}
\includegraphics[angle=0,scale=0.58]{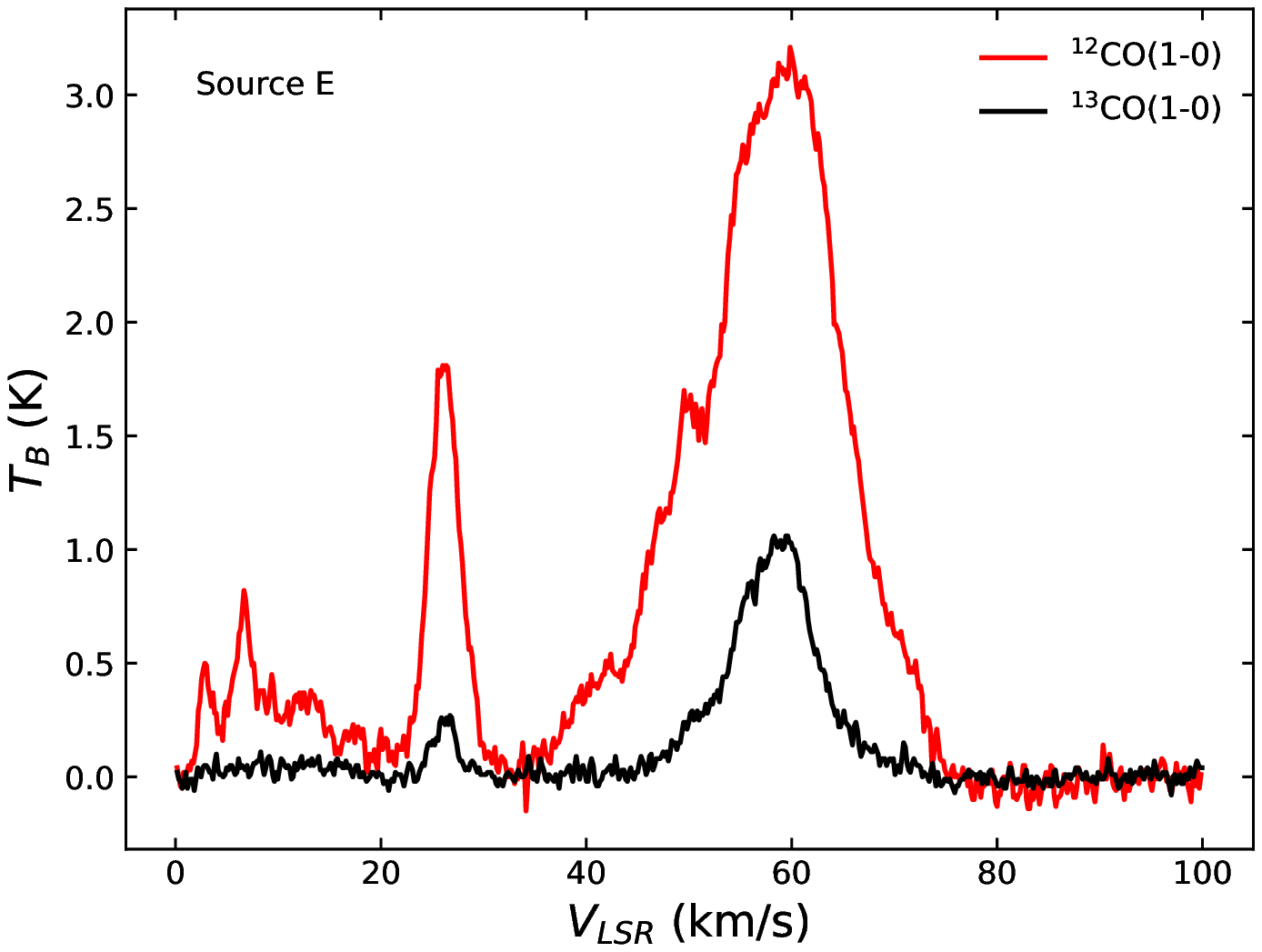}
\includegraphics[angle=0,scale=0.58]{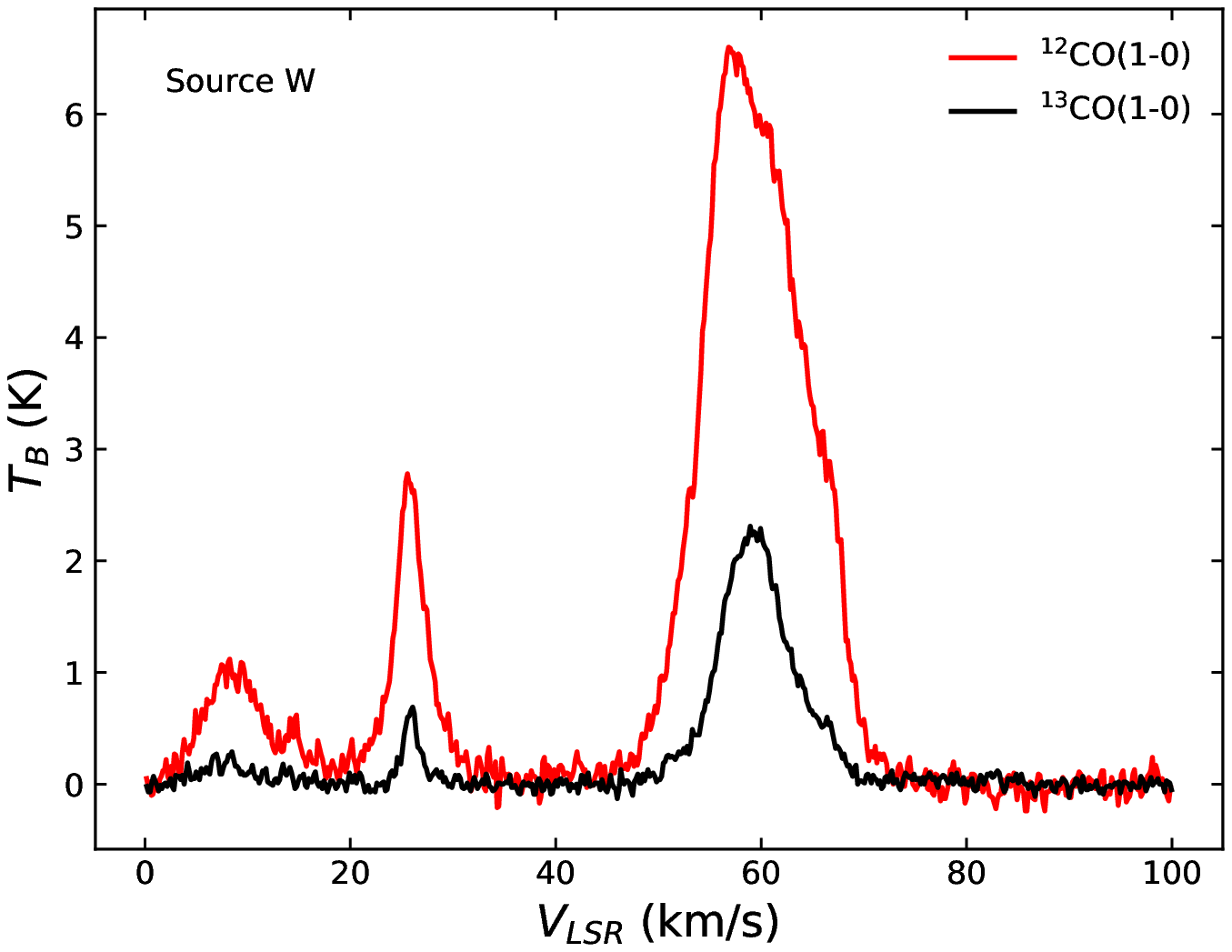}
\caption{$^{12}$CO($J$=1--0; red) and $^{13}$CO($J$=1--0; black) spectra of molecular gas toward Source E and Source W. These spectra are extracted from regions of $0.25\degr \times 0.25\degr$ and $0.10\degr \times 0.10\degr$ around the centre of Source E and Source W, respectively.}
\label{COsed}
\end{figure*}

\begin{figure*}
\includegraphics[angle=0,scale=0.37]{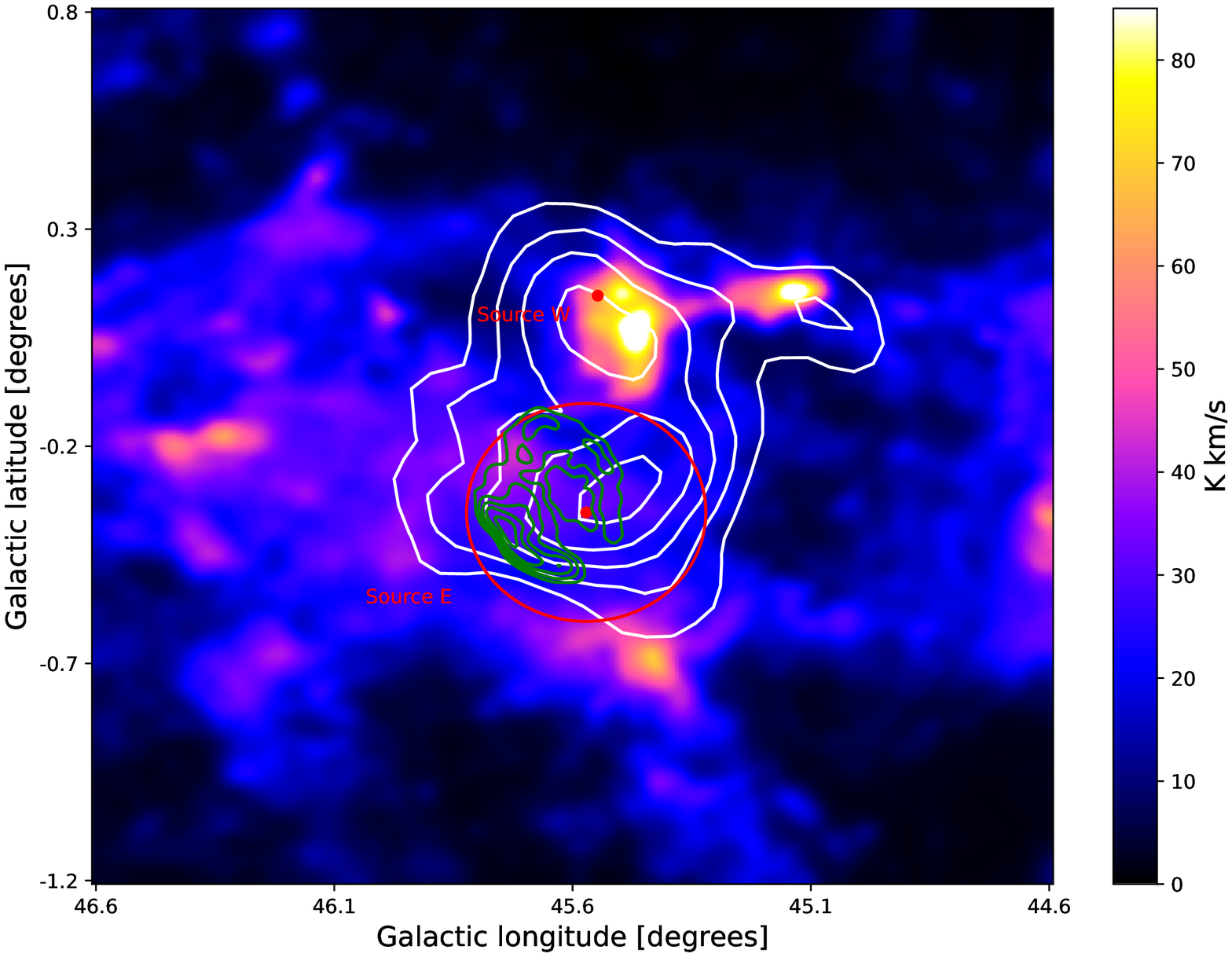}
\includegraphics[angle=0,scale=0.37]{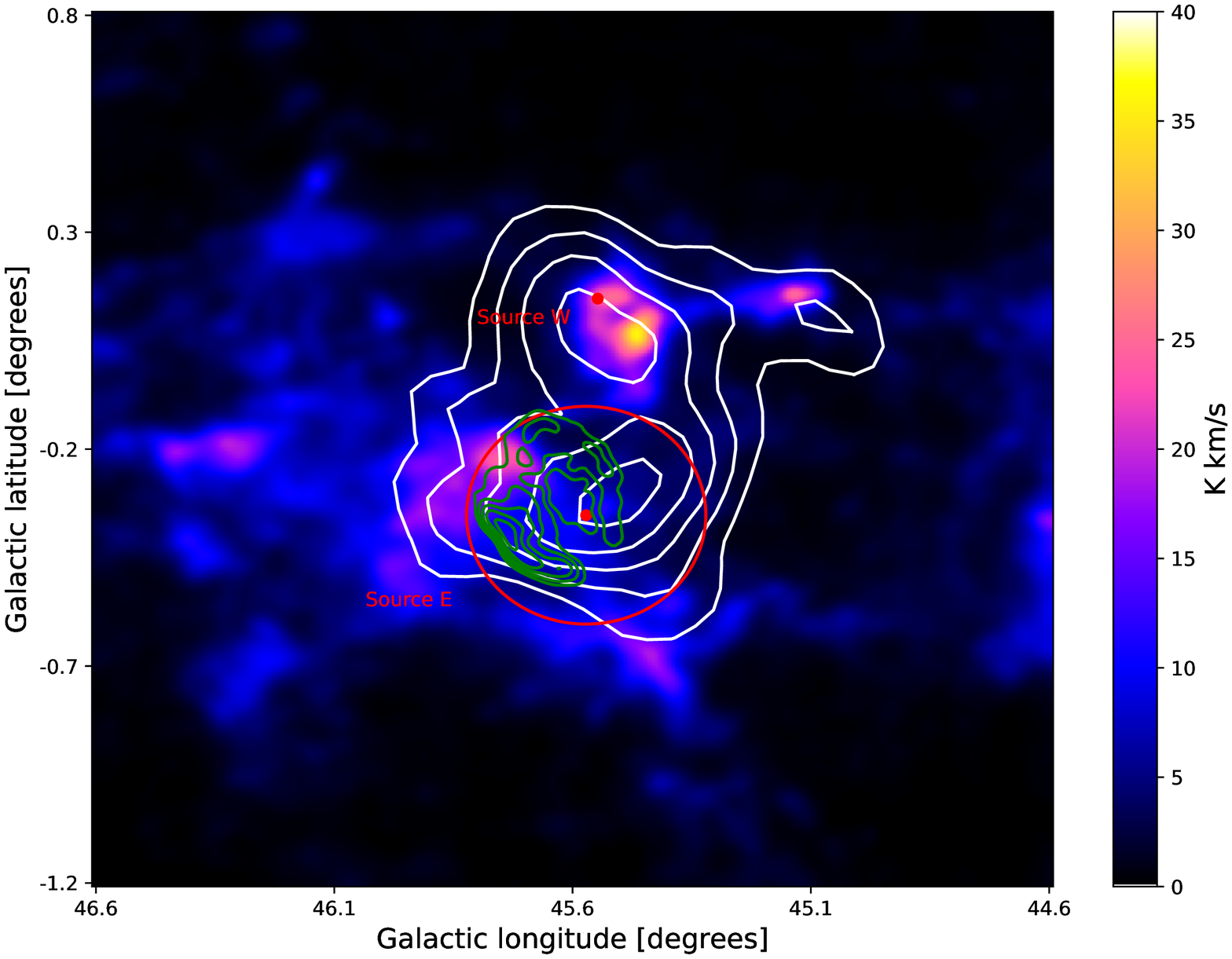}
\caption{Integrated $^{12}$CO (left plane) and $^{13}$CO (right plane) emission toward SNR G45.7-00.4 in the interval of $\rm 52.0-62.0\ km \ s^{-1}$.
The green contours show the radio emission at 200 MHz as shown in Figure \ref{rxmap}.
The white significance contours correspond to $\sqrt{\rm TS}$ at 4.0, 5.0, 6.0, 7.0 and 8.0 as shown in Figure \ref{Gevmap1}.
The red circle shows the size of Source E, the red points show the centre of Source E and Source W.
}
\label{COmap}
\end{figure*}

We make use of the data from the Milky Way Imaging Scroll Painting (MWISP\footnote{\url{http://english.dlh.pmo.cas.cn/ic/}}) project, which is a multi-line survey in $^{12}$CO/$^{13}$CO/C$^{18}$O along the northern galactic plane with PMO-13.7m telescope. The detailed observing strategy, the instrument, and the quality of the CO observations can be found in \citet{Su2019}. In this section, we present the results of the MWISP CO survey for $2\degr \times 2\degr$ regions centre of $(l,b)=(45.50\degr,0.30\degr)$.
Broad profiles of $^{12}$CO emission usually come from the turbulent molecular gas that is readily and significantly influenced by local shocks. However, the $^{13}$CO emission, usually optically thin, arises in the quiescent dense gas along the whole line of sight (LOS).
We have inspected the $^{12}$CO and $^{13}$CO line profiles of the MCs toward the Source E and Source W region, to search for kinematic evidence for gas distribution due to external interaction\citep{Frail1998,Reach2005,Jiang2010,Zhou2011,Kilpatrick2016,Liu2020}.
We do not find any significant evidence of broadenings or asymmetries in $^{12}$CO line with respect to the narrow $^{13}$CO line.
The spectra of $^{12}$CO and $^{13}$CO emission toward Source E and Source W are shown in Figure \ref{COsed}. We find that the spectra of $^{12}$CO emission shows multi-peaks, but only two peaks shown in the spectra of $^{13}$CO emission. Considering that the $^{12}$CO emission at a systemic velocity of 0--40 km/s is widely distributed toward the inner Galaxy \citep{Su2019}, and the peak of $^{13}$CO emission at a velocity of $\sim 25 \rm \ km/s$ is too weak with respect to the peak at a velocity of $\sim 58 \rm \ km/s$, we think that the peak of $\sim 25 \rm \ km/s$ can be ignored.
The molecular gas complex at a systemic velocity of $\rm V_{LSR} \sim +58 \ km \ s^{-1}$ in the direction of the GeV source corresponds to a near distance $\rm d_{near}=3.7$ kpc and far distance $\rm d_{far}=7.4$ kpc \citep{Brand1993,Reid2019}.
In the following, we parameterize the distance of the molecular gas complex as $d=3.7 d_{3.7}{\rm kpc}$.
Figure \ref{COmap} shows integrated CO emission map concentrated toward SNR G45.7-00.4 in the velocity interval of $\rm 52.0-62.0\ km \ s^{-1} $.
Apparently, there is good correlation between the gas distribution and the intensity of $\gamma$-ray emission. Particularly, the point-like GeV Source W coincides well with the densest region of the gas distribution.
Adopting the mean CO-to-$\rm H_{2}$ mass conversion factor $\rm X_{CO}=2\times10^{20} \ cm^{-2} \ K^{-1} \ km^{-1} \ s$ \citep{Bolatto2013},
we estimate that the total mass of gas within $0.25\degr$ of the Source E is about $1.5\times 10^{5}d_{3.7}^{2} \ M_{\odot}$.
Assuming a spherical geometry of the gas distribution, the average $\rm H_2$ cubic density in this region is about $\rm 280d_{3.7}^{-1} \ cm^{-3}$.
The densest CO molecular gas that coincides with Source W corresponds to a molecular cloud identified by \citet{Miv2017}, which is named ``[MML2017] 683'' in SIMBAD \footnote{\url{http://simbad.u-strasbg.fr/simbad/sim-ref?querymethod=bib&simbo=on&submit=submit+bibcode&bibcode=2017ApJ...834...57M}}.  \citet{Miv2017} obtained a velocity of $\rm V_{LSR}= 57.81\pm5.43 \, km\, s^{-1}$, consistent with our measurement.Since Source W is a point-like source, we  estimate the mass and density of the gas by considering the CO emission within $0.1\degr$ of the Source W. We find that the  mass of gas within $0.1\degr$ of the Source W is about $6.2\times 10^{4}d_{3.7}^{2} \ M_{\odot}$ and the average $\rm H_2$ cubic density is about $\rm 2000d_{3.7}^{-1} \, cm^{-3}$.



\section{Interpretation of the GeV emission}

\subsection{Source E}

The morphological  analysis shows  that the extended gamma-ray source (Source E) is coincident with the radio SNR, suggesting that it  represents the GeV emission of the shell of SNR G045.7-00.4. 


Assuming a real association between the  molecular cloud and SNR G045.7-00.4, the distance of SNR G045.7-00.4 is about $d=3.7d_{3.7}{\rm \ kpc}$. We can also estimate the distance of the SNR using the empirical radio surface-brightness-to-diameter ($\Sigma–D$) relations for SNRs. The radio surface-brightness of SNR G045.7-00.4 is $1.2\times10^{-21}{\rm W \ m^{-2} \ Hz^{-1} \ Sr^{-1}}$, suggesting a diameter of about $30 {\rm pc}$ \citep{Pavlovic2013}. Note that the diameter of the SNR inferred from the $\Sigma–D$ relation has a large uncertainty. Give the angular radius of $\theta\simeq 11'$ for the radio SNR, the distance is inferred to be about $4.3{\rm \ kpc}$.  In the following, we adopt the distance of $d=3.7{\rm \ kpc}$ as a reference value for SNR G045.7-00.4.  The luminosities of the gamma-ray emission in 0.1-500 GeV of Source E is $(6.60\pm1.27)\times10^{34} d_{3.7}^2{\rm \, erg \, s^{-1}}$.

The radius of the radio SNR is about $r=\theta d \simeq 12\,$pc. For an SNR in the Sedov-Taylor expansion phase, the size of the SNR grows with time as 
$r=2.3 {\rm pc} (W_k/10^{51}{\rm erg})^{1/5} (n/1{\rm cm^{-3}})^{-1/5}(t/100{\rm yr})^{2/5}$, where $n$ is the number density of the ambient medium and $W_k$ is the kinetic energy of the supernova explosion. 
Then, we can estimate the age of SNR G045.7-00.4 from its radius, i.e., $t\simeq 7\times 10^3{\rm yr}  (W_k/10^{51}{\rm erg})^{-1/2}(n/1{\rm cm^{-3}})^{1/2}$. {It is not clear when the SNR encountered the molecular cloud. The SNR could experience most of the evolution time so far in ISM and encountered molecular cloud just recently, or, alternatively, the SNR could have already expanded in the molecular gas since very beginning.} For a density of $n=1-280\,{\rm cm^{-3}}$, the age of the SNR could span in a large range of about $(0.7-10)\times10^4{\rm yr}$. 

\begin{figure*}
\includegraphics[angle=0,scale=0.38]{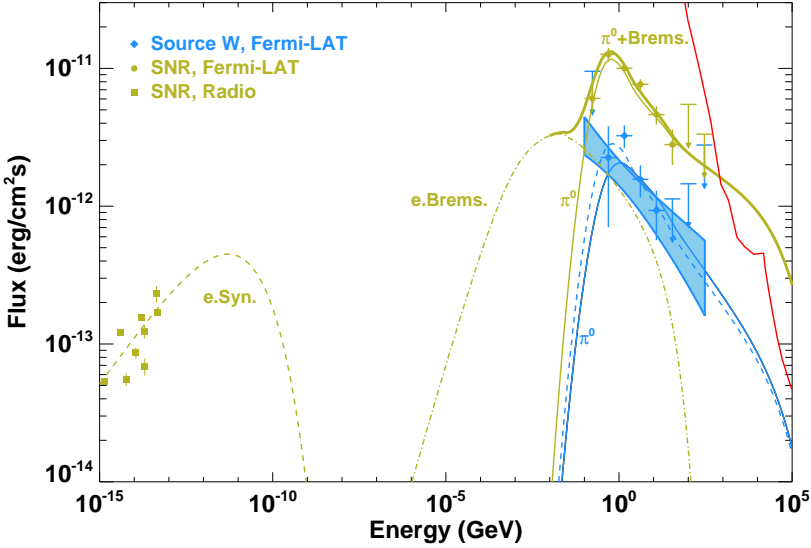}
\caption {Modeling of the multi-wavelength SED of SNR G045.7-00.4 with the leptonic and hadronic model.
The radio spectra data are derived from CATS database \citep{Verkhodanov2005}.
The shaded blue regions are the 68\% confidence range of the LAT spectra of Source W, taking into account the statistical errors only.
The error bars of the \emph{Fermi}-LAT spectra are represent statistical uncertainties.
Model parameters for SNR G045.7-00.4 are $B=80\,\mu$G, $W_{\rm e}=1\times10^{48} \ $erg, $E_{e,\rm max}=50\,$GeV, $s_e=2.3$, $W_{\rm p,SNR}=2\times10^{48} \ $erg, $ E_{p, \rm max}=1 \ $PeV, $s_p=2.3$ and $n=280{\rm cm^{-3}}$. For Source W, the model parameters are $\eta_p=0.06$, $\chi=1$ for the case of $t_{\rm age}=7000\,$yr (blue solid curve) and $\eta_p=0.1$, $\chi=0.1$ for the case of $t_{\rm age}=10^5\,$yr (blue dashed curve). The target gas density is $n=2000 {\rm cm^{-3}}$. See text for more details.
The solid red curve shows the LHAASO point-source sensitivity of one-year exposure \citep{Bai2019}.
}
\label{SED}
\end{figure*}

As source E  coincide spatially with dense molecular clouds, a natural mechanism for the emission is hadronic emission produced by cosmic rays interacting with the gas in the molecular clouds. Therefore, we model the GeV spectra of both sources with the hadronic model, which are shown in Figure \ref{SED}. 
The GeV emission is dominated by $\pi^0$-decay process.  Below 1\,GeV, the $\pi^0$ bump is slightly steeper than the data, and we speculate that the bremsstrahlung emission of electrons may have a sub-dominant contribution to the flux. In the meanwhile, the radio emission is produced by the same electrons via the synchrotron radiation. Combined with the $<1\,$GeV data and the radio data, we find that a magnetic field of $80{\rm \mu G}$ gives a reasonable fitting. 
Note that, we here take the matter density of $n=280{\rm cm^{-3}}$ and the magnetic field  depends on the matter density.
The electron spectrum is assumed to follow the form $E_e^{-s_e}\exp(-E_e/E_{\rm e,max})$ with $s_e=2.3$ and we use an arbitrary value of $E_{\rm e, max}=50\,$GeV for the maximum electron energy. The total electron energy above 5\,MeV is taken to be $W_e=10^{48}\,$erg. We note that the electron spectrum may probably be a broken power-law due to the cooling effect. However, lacking of optical and X-ray data, the parameters of the broken power-law function cannot be constrained. Thus, we simply assume a single power-law function with an exponential cutoff at $E_{\rm e,max}$, which is also arbitrarily set. The proton spectrum is also assumed to be a power-law function with a high-energy cutoff. The proton spectrum slope is assumed to be $s_p=2.3$ and the total energy of cosmic rays is about $W_{\rm p,SNR}=2\times 10^{48}{\rm \ erg}$ above 1\,GeV.

\subsection{Source W}

Besides SNR G045.7-00.4, there are a few HII regions and pulsars in the GeV emission region (see Figure \ref{Gevmap1}).  The brightest HII region is GAL045.47+00.07, which is a compact HII region and locates at a distance of $7.7\pm0.6{\rm \ kpc}$ \citep{Kolpak2003}.  At such a distance, the luminosity of Source W would be $7.9\times10^{34} \rm \, erg\, s^{-1}$ if it is associated with GAL 045.47+00.07. An infrared source IRAS 19124+1106 appears projected in close association with  GAL 045.47+00.07. The derived IRAS luminosity is $3.3\times10^4 L_\odot$, correspond to an O9.5 ZAMS star \citep{Rodriguez1998}. The kinetic energy luminosity in the stellar wind  of such a star is about $10^{35}{\rm erg \, s^{-1}}$.  Unless the efficiency of converting the wind power to CRs is close to $100\%$, which seems unlikely, the wind power is insufficient to explain the gamma-ray luminosity.

Pulsars or associated PWNe could also be potential GeV gamma-ray sources. 
We search for pulsars in the GeV region according to the ATNF (Australia Telescope National Facility\footnote{\url{https://www.atnf.csiro.au/research/pulsar/psrcat/}}) Pulsar Catalogue \citep{Manchester2005}. We find that only one pulsar (PSR J1914+1122, 0.121$\degr$ distant from centre of Source W) locating within $3\sigma$ error circle of the Source W. But it is too weak to power the $\gamma$-ray emission, since its spin-down luminosity is only $1.2 \times 10^{32} \rm \, erg \, s^{-1}$. 

Active galactic nucleus (AGNs) are a known $\gamma$-ray-emitting source class. Therefore, we search the possible AGNs counterparts within $0.2\degr$ of the centre of Source W in the SIMBAD. But we do not find any known AGNs counterparts.

Source W  is spatially coincident with the densest region  of the molecular cloud. This is reminiscent of  SNR W28, where a nearby GeV source is located at  the  boundary of W28. Like W28, Source W can be interpreted as arising from escaping cosmic rays that interact with surrounding molecular cloud. The luminosity of the gamma-ray emission in 0.1-500 GeV of  Source W is  $(1.81\pm0.36)\times10^{34}d_{3.7}^2{\rm \ erg \ s^{-1}}$.

For Source W, we assume that the GeV emission is also dominated by $\pi^0$-decay process. Note that, given $d=3.7\,$kpc, the radius of the dense molecular cloud around Source W is about $0.1\degr$ or $R_c=6.5\,$pc, and Source W lies at a distance of $0.46\degr$ from the center of G045.7-00.4 or $r=30\,$pc. Let us assume that cosmic rays escape isotropically from the sphere of the shock. Following previous studies \citep{Gabici2009, Ohira11}, we assume the escape of cosmic rays of energy $E_p$ starts at $t_{\rm esc}(E_p)=t_{\rm ST}(E/E_{p, \rm max})^{-5/2\delta}$ with $\delta$ being a model parameter. $t_{\rm ST}$ is the staring time of the Sedov-Taylor phase, which can be approximately given by  $t_{\rm ST}\approx 400 (W_k/10^{51}{\rm erg})^{-1/2}(M_{\rm ej}/2M_\odot)^{5/6}(n/1{\rm cm^{-3}})^{-1/3}\,$yr \citep{Truelove1999}. Here  $M_{\rm ej}$ is the mass of the ejecta. This corresponds to an escape radius $R_{\rm esc}=R_{\rm ST}(E_p/E_{p, \rm max})^{\delta}$ with $R_{\rm ST}\approx 3(M_{\rm ej}/2M_\odot)^{1/3}(n/1{\rm cm^{-3}})^{-1/3}\,$pc. The differential density of escaping CRs at the position of Source W reads \citep{Ohira11, Celli19}
\begin{equation}
\begin{split}
    &n_{\rm CR}(E_p,r)=\frac{N_{p,\rm esc}(E_p)}{4\pi^{3/2}R_{\rm esc}R_{\rm diff}r}\times\\
    &\left\{\exp\left[-\left(\frac{R_{\rm esc}-r}{R_{\rm diff}}\right)^2\right]-\exp\left[-\left(\frac{R_{\rm esc}+r}{R_{\rm diff}}\right)^2\right]\right\},
\end{split}    
\end{equation}
where $N_{p,\rm esc}(E_p)$ is the spectrum of escaping proton, which is assumed to be $N_{p,\rm esc}(E_p)=\eta_pW_kE_p^{-2}/\ln(E_{p,\rm max}/E_{p,\rm min})$. $\eta_p$ is defined as the fraction of the kinetic energy converted into runaway CR energy.   $R_{\rm diff}=2\sqrt{D(E_p)t}$ where $D(E_p)$ is the diffusion coefficient around the SNR. We assume a homogeneous diffusion coefficient and paramterize it as $D(E_p)=\chi D_{\rm ISM}=\chi 10^{28}(E_p/10\,{\rm GeV})^{1/3}\,{\rm cm^2s^{-1}}$ with $\chi$ being the ratio between the diffusion coefficient to the average one of the Galactic plane. Then, we can calculate the expected gamma-ray flux of Source W by
\begin{equation}
    \frac{dN}{dE_\gamma}=\frac{n_wc}{4\pi d^2}\left(\frac{4}{3}\pi R_c^3\right)\int \frac{d\sigma_{pp}}{dE_\gamma}\left(E_\gamma,E_p\right)n_{\rm CR}(E_p,r)dE_p 
\end{equation}
where $d\sigma_{pp}/dE_\gamma$ is the differential $pp$ inelastic cross section for gamma-ray production \citep{Kafexhiu14} and $(4/3)\pi R_c^3$ accounts for the approximate volume of the molecular cloud at the location of Source W.

Now let us consider two limiting scenarios. In one scenario, the SNR expanded in typical ISM of density $1\,\rm cm^{-3}$ during most of the evolution time leading to an age of 7000\,yr for the SNR, while, in the other, the SNR has expanded in the molecular cloud of density $280\,cm^{-3}$ since long time ago, leading to an age of $10^5\,$yr. In Fig.~\ref{SED}, we also show the resulting pionic gamma-ray flux produced by runaway CRs for Source W in the two limiting scenarios. The solid blue curve represents the scenario of $t_{\rm age}=7000\,$yr with $\eta_p=0.06$, $\delta=15$ and $\chi=1$, while the dashed blue curve represents the scenario of $t_{\rm age}=10^5\,$yr with $\eta_p=0.1$, $\delta=6.5$ and $\chi=0.1$. Both scenarios can give acceptable fitting to the \textit{Fermi}-LAT data, but we note that the value of $\delta$ required in the former scenario is quite large. This is because given an age of only 7000\,yr for the SNR, we need the escape CR energy quickly drop to 10\,GeV in order to account for the measured 1\,GeV flux,  otherwise the 10\,GeV protons would not have sufficient time to arrive at the position of Source W. Assuming a larger diffusion coefficient could alleviate this problem, but in the mean while requires an unreasonably large energy budget for the runaway CRs. In the latter scenario, we simply employ $\chi=0.1$, following \citet{Ohira11}.  With $\eta_p=0.06-0.1$, the  total energy of  CRs escaping from the SNR is $\sim (0.6-1)\times10^{50}{\rm erg}$, assuming the escaping is isotropic. We also show the Large High Altitude Air Shower Observatory (LHAASO) point-source sensitivity of one-year exposure in the plot \citep{Bai2019}, which indicates that G045.7-00.4 could be detectable by LHAASO as long as the power-law spectrum of CR protons extends to a sufficiently high energy (e.g. $E_{\rm p,max}=1\,$PeV as assumed in the calculation).
Note that the angular resolution of LHAASO is about $0.3\degr$ at energies above  10 TeV \citep{Bai2019}. Due to the poor angular resolution of LHAASO, it may be difficult to distinguish between the emission from Source E and Source W.


\section{Summary }

We analyzed the GeV $\gamma$-ray emission in the vicinity of SNR G045.7-00.4 using 12 years of {\rm Fermi}-LAT data. We detected one extended gamma-ray source coincident with the radio SNR and a point-like source located west of  the radio boundary of the SNR.  Both gamma-ray sources overlap with the molecular clouds  in the velocity range of 52 to 62 ${\rm km s^{-1}}$. The gamma-ray emission from both sources can be interpreted as arsing from the decay of $\pi^0$s produced by the interaction of CRs with the dense molecular gas. In particular, the western source may result from the escaping CRs from G045.7-00.4 during an earlier epoch.  The GeV  data of  Source W can be consistently explained by this scenario assuming that the total energy in escaping CRs is about $10^{50}{\rm erg}$.  Depending on the  age of the SNR, the diffusion coefficient of CRs at energy ${\rm 10 \ GeV }$ is estimated to be $ (1-10)\times 10^{27}{\rm cm^2 \ s^{-1}}$.

Interestingly, X-ray observations by ROSAT PSPC\footnote{\url{https://skyview.gsfc.nasa.gov/current/cgi/runquery.pl}} (0.1 to 2.4 keV) shows an extremely bright source, GRS 1915+105, located only $0.25\degr$ and $0.38\degr$ away from the centre of Source E and Source W, respectively.  GRS 1915+105 is  a superluminal microquasar  at a distance of $8.6_{-1.6}^{+2.0}$ kpc \citep{Reid2014},  with a black hole accreting matter from a low-mass star. The X-ray luminosity of the microquasar is as high as a few $10^{38}{\rm erg \ s^{-1}}$. The energy transport rate of the jets has so far only been estimated during times when milliarcseconds jet components are detectable in the radio. It is estimated that  the energy transport rate during such periods is between several $10^{37} {\rm erg \ s^{-1}}$ and almost
$10^{42} {\rm erg \ s^{-1}}$, depending on the composition of the jet material
and its velocity  \citep{Fender1999,Gliozzi1999,Fender2000}.  However,  for the extended GeV emission that we consider here, only the energy transport rate of
the jets averaged over timescales comparable to the formation timescale of the large-scale structure is important. It was suggested that   two IRAS regions, IRAS 19124+1106 and IRAS 19132+1035, are the  impact regions of large-scale jets from GRS 1915+105 \citep{Rodriguez1998,Kaiser2004}, as the two
impact regions are almost perfectly symmetric about the
position of GRS 1915+105 and  lie in the directions that
the small-scale jets of this source point to. Modeling of the 
the  dynamics and radiation of the possible large-scale jets gives a time-averaged energy transportation rate $>10^{36}{\rm erg \ s^{-1}}$ \citep{Kaiser2004} .  So the jets from GRS 1915+105 could, in principle, contribute at least partly to the GeV emission.

\acknowledgments
We thank Felix Aharonian, Siming Liu and Jian Li for invaluable discussions. This work was supported by the National KeyR \& D program of China under the grant 2018YFA0404203 and the NSFC grants 11625312 and  U2031105.

This research made use of the data from the Milky Way Imaging Scroll Painting (MWISP) project, which is a multi-line survey in $^{12}$CO/$^{13}$CO/C$^{18}$O along the northern galactic plane with PMO-13.7m telescope. We are grateful to all the members of the MWISP working group, particularly the staff members at PMO-13.7m telescope, for their long-term support. MWISP was sponsored by National Key R\&D Program of China with grant 2017YFA0402701 and CAS Key Research Program of Frontier Sciences with grant QYZDJ-SSW-SLH047.

\begin{table}[ht!]
\caption{Morphological models tested for the GeV gamma-ray emission above 3 GeV.}
\begin{center}
    \begin{tabular}{lcccc}
        \hline
        Morphology ($>$ 3 GeV) & Extension & TS & $\rm TS_{ext}$ & $N_{\rm dof}$\\ \hline
        PS+PS           & --       & 110.18 & -- & 8 \\
         Disk+PS    & $0.25\degr_{-0.05\degr}^{+0.08\degr}$       	   &  135.98 &  25.81 &9\\
        Gaussian+PS   & {\bf$0.24\degr_{-0.06\degr}^{+0.07\degr}$ }         & 139.30 & 29.13&9\\
         \hline
    \end{tabular}
    \end{center}
    \tablenotetext{}{{\bf Note:} The morphological models PS, Disk and Gaussian represent a point-like source model, an uniform disk model and a two-dimensional Gaussian model, respectively. Extension represents the radius containing 68\% of the intensity of  Source E for the tested models.  Source W does not show significant extension, so it is treated as a point-like source. Assuming  a disk shape for  Source W, the upper limit extension is 21 arcmin at the 95\% confidence level. The TS is evaluated from the likelihood ratio between two models with and without the morphological model of interest. The $\rm TS_{ext}$ corresponds to the extension significance of the Source E, since the Source W is not find significant extension and treated as a point-like source. $N_{\rm dof}$ represents the number of degrees of freedom for each model.}
    \label{tab:morph}
\end{table}{}

\begin{thebibliography}{}

\bibitem[Abdo et al.(2010a)]{Abdo2010} Abdo, A.~A., Ackermann, M., Ajello, M., et al.\ 2010, \apj, 718, 348. doi:10.1088/0004-637X/718/1/348


\bibitem[Abdollahi et al.(2020a)]{Abdollahi2020a} Abdollahi, S., Acero, F., Ackermann, M., et al.\ 2020, \apjs, 247, 33. doi:10.3847/1538-4365/ab6bcb
\bibitem[Abdollahi et al.(2020b)]{Abdollahi2020b} Abdollahi, S., Ballet, J., Fukazawa, Y., et al.\ 2020, \apj, 896, 76. doi:10.3847/1538-4357/ab91b3


\bibitem[Acero et al.(2016)]{Acero2016} Acero, F., Ackermann, M., Ajello, M., et al.\ 2016, \apjs, 224, 8. doi:10.3847/0067-0049/224/1/8

\bibitem[Aharonian \& Atoyan(1996)]{Aharonian1996} Aharonian, F.~A. \& Atoyan, A.~M.\ 1996, \aap, 309, 917
\bibitem[Aharonian et al.(2004)]{Aharonian2004} Aharonian, F.~A., Akhperjanian, A.~G., Aye, K.-M., et al.\ 2004, \nat, 432, 75. doi:10.1038/nature02960
\bibitem[Aharonian et al.(2008)]{Aharonian2008a} Aharonian, F., Akhperjanian, A.~G., Bazer-Bachi, A.~R., et al.\ 2008, \aap, 481, 401. doi:10.1051/0004-6361:20077765

\bibitem[Atwood et al.(2009)]{Atwood2009} Atwood, W.~B., Abdo, A.~A., Ackermann, M., et al.\ 2009, \apj, 697, 1071. doi:10.1088/0004-637X/697/2/1071


\bibitem[Bai et al.(2019)]{Bai2019} Bai, X., Bi, B.~Y., Bi, X.~J., et al.\ 2019, arXiv:1905.02773
\bibitem[Bayandina et al.(2015)]{Bayandina2015} Bayandina, O.~S., Val'tts, I.~E., \& Kurtz, S.~E.\ 2015, Astronomy Reports, 59, 998. doi:10.1134/S1063772915110025
\bibitem[Brand \& Blitz(1993)]{Brand1993} Brand, J. \& Blitz, L.\ 1993, \aap, 275, 67
\bibitem[Bolatto et al.(2013)]{Bolatto2013} Bolatto, A.~D., Wolfire, M., \& Leroy, A.~K.\ 2013, \araa, 51, 207. doi:10.1146/annurev-astro-082812-140944

\bibitem[Celli et al.(2019)]{Celli19} Celli, S., Morlino, G., Gabici, S., et al.\ 2019, \mnras, 490, 4317. doi:10.1093/mnras/stz2897

\bibitem[Frail \& Mitchell(1998)]{Frail1998} Frail, D.~A. \& Mitchell, G.~F.\ 1998, \apj, 508, 690. doi:10.1086/306452
\bibitem[Fender et al.(1999)]{Fender1999} Fender, R.~P., Garrington, S.~T., McKay, D.~J., et al.\ 1999, \mnras, 304, 865. doi:10.1046/j.1365-8711.1999.02364.x
\bibitem[Fender \& Pooley(2000)]{Fender2000} Fender, R.~P. \& Pooley, G.~G.\ 2000, \mnras, 318, L1. doi:10.1046/j.1365-8711.2000.03847.x
\bibitem[Fuerst et al.(1987)]{Fuerst1987} Fuerst, E., Reich, W., Reich, P., et al.\ 1987, \aaps, 69, 403


\bibitem[Gabici et al.(2009)]{Gabici2009} Gabici, S., Aharonian, F.~A., \& Casanova, S.\ 2009, \mnras, 396, 1629. doi:10.1111/j.1365-2966.2009.14832.x

\bibitem[Gliozzi et al.(1999)]{Gliozzi1999} Gliozzi, M., Bodo, G., \& Ghisellini, G.\ 1999, \mnras, 303, L37. doi:10.1046/j.1365-8711.1999.02436.x
\bibitem[Green(2019)]{Green2019} Green, D.~A.\ 2019, Journal of Astrophysics and Astronomy, 40, 36. doi:10.1007/s12036-019-9601-6


\bibitem[Hanabata et al.(2014)]{Hanabata2014} Hanabata, Y., Katagiri, H., Hewitt, J.~W., et al.\ 2014, \apj, 786, 145. doi:10.1088/0004-637X/786/2/145

\bibitem[Helene(1983)]{Helene1983} Helene, O.\ 1983, Nuclear Instruments and Methods in Physics Research, 212, 319. doi:10.1016/0167-5087(83)90709-3


\bibitem[Hurley-Walker et al.(2019)]{Hurley2019} Hurley-Walker, N., Hancock, P.~J., Franzen, T.~M.~O., et al.\ 2019, \pasa, 36, e047. doi:10.1017/pasa.2019.37

\bibitem[Jiang et al.(2010)]{Jiang2010} Jiang, B., Chen, Y., Wang, J., et al.\ 2010, \apj, 712, 1147. doi:10.1088/0004-637X/712/2/1147

\bibitem[Kafexhiu et al.(2014)]{Kafexhiu14} Kafexhiu, E., Aharonian, F., Taylor, A.~M., et al.\ 2014, \prd, 90, 123014. doi:10.1103/PhysRevD.90.123014


\bibitem[Kaiser et al.(2004)]{Kaiser2004} Kaiser, C.~R., Gunn, K.~F., Brocksopp, C., et al.\ 2004, \apj, 612, 332. doi:10.1086/422466
\bibitem[Kilpatrick et al.(2016)]{Kilpatrick2016} Kilpatrick, C.~D., Bieging, J.~H., \& Rieke, G.~H.\ 2016, \apj, 816, 1. doi:10.3847/0004-637X/816/1/1

\bibitem[Kolpak et al.(2003)]{Kolpak2003} Kolpak, M.~A., Jackson, J.~M., Bania, T.~M., et al.\ 2003, \apj, 582, 756. doi:10.1086/344752

\bibitem[Liu et al.(2020)]{Liu2020} Liu, Q.-C., Chen, Y., Zhou, P., et al.\ 2020, \apj, 892, 143. doi:10.3847/1538-4357/ab7a22




\bibitem[Malkov \& Drury(2001)]{Malkov2001} Malkov, M.~A. \& Drury, L.~O.\ 2001, Reports on Progress in Physics, 64, 429. doi:10.1088/0034-4885/64/4/201
\bibitem[Manchester et al.(2005)]{Manchester2005} Manchester, R.~N., Hobbs, G.~B., Teoh, A., et al.\ 2005, \aj, 129, 1993. doi:10.1086/428488


\bibitem[Miville-Desch{\^e}nes et al.(2017)]{Miv2017} Miville-Desch{\^e}nes, M.-A., Murray, N., \& Lee, E.~J.\ 2017, \apj, 834, 57. doi:10.3847/1538-4357/834/1/57


\bibitem[Ohira et al.(2011)]{Ohira11} Ohira, Y., Murase, K., \& Yamazaki, R.\ 2011, \mnras, 410, 1577. doi:10.1111/j.1365-2966.2010.17539.x


\bibitem[Pavlovi{\'c} et al.(2013)]{Pavlovic2013} Pavlovi{\'c}, M.~Z., Uro{\v{s}}evi{\'c}, D., Vukoti{\'c}, B., et al.\ 2013, \apjs, 204, 4. doi:10.1088/0067-0049/204/1/4
\bibitem[Peron et al.(2020)]{Peron2020} Peron, G., Aharonian, F., Casanova, S., et al.\ 2020, \apjl, 896, L23. doi:10.3847/2041-8213/ab93d1

\bibitem[Reach et al.(2005)]{Reach2005} Reach, W.~T., Rho, J., \& Jarrett, T.~H.\ 2005, \apj, 618, 297. doi:10.1086/425855
\bibitem[Reid et al.(2014)]{Reid2014} Reid, M.~J., McClintock, J.~E., Steiner, J.~F., et al.\ 2014, \apj, 796, 2. doi:10.1088/0004-637X/796/1/2
\bibitem[Reid et al.(2019)]{Reid2019} Reid, M.~J., Menten, K.~M., Brunthaler, A., et al.\ 2019, \apj, 885, 131. doi:10.3847/1538-4357/ab4a11

\bibitem[Rodriguez Marrero et al.(2008)]{Rodriguez Marrero2008} Rodriguez Marrero, A.~Y., Torres, D.~F., de Cea del Pozo, E., et al.\ 2008, \apj, 689, 213. doi:10.1086/592562
\bibitem[Rodriguez \& Mirabel(1998)]{Rodriguez1998} Rodriguez, L.~F. \& Mirabel, I.~F.\ 1998, \aap, 340, L47


\bibitem[Su et al.(2019)]{Su2019} Su, Y., Yang, J., Zhang, S., et al.\ 2019, \apjs, 240, 9. doi:10.3847/1538-4365/aaf1c8

\bibitem[Truelove \& McKee(1999)]{Truelove1999} Truelove, J.~K. \& McKee, C.~F.\ 1999, \apjs, 120, 299. doi:10.1086/313176


\bibitem[Uchiyama et al.(2012)]{Uchiyama2012} Uchiyama, Y., Funk, S., Katagiri, H., et al.\ 2012, \apjl, 749, L35. doi:10.1088/2041-8205/749/2/L35

\bibitem[Verkhodanov et al.(2005)]{Verkhodanov2005} Verkhodanov, O.~V., Trushkin, S.~A., Andernach, H., et al.\ 2005, Bulletin of the Special Astrophysics Observatory, 58, 118

\bibitem[Zhou \& Chen(2011)]{Zhou2011} Zhou, P. \& Chen, Y.\ 2011, \apj, 743, 4. doi:10.1088/0004-637X/743/1/4



\end{thebibliography}
\end{document}